\documentclass[11pt, epsf]{article}

\usepackage{epsfig}
\usepackage{amssymb}
\usepackage{graphicx}
\usepackage{color}
\usepackage{subfigure}
\usepackage{dsfont}

\begin{document}

\baselineskip 6mm
\renewcommand{\thefootnote}{\fnsymbol{footnote}}


\newcommand{\nc}{\newcommand}
\newcommand{\rnc}{\renewcommand}



\newcommand{\tcb}{\textcolor{blue}}
\newcommand{\tcr}{\textcolor{red}}
\newcommand{\tcg}{\textcolor{green}}


\def\be{\begin{equation}}
\def\ee{\end{equation}}
\def\ba{\begin{array}}
\def\ea{\end{array}}
\def\bea{\begin{eqnarray}}
\def\eea{\end{eqnarray}}
\def\nn{\nonumber\\}


\def\ct{\cite}
\def\la{\label}
\def\eq#1{(\ref{#1})}


\def\a{\alpha}
\def\b{\beta}
\def\g{\gamma}
\def\G{\Gamma}
\def\d{\delta}
\def\D{\Delta}
\def\e{\epsilon}
\def\et{\eta}
\def\ph{\phi}
\def\Ph{\Phi}
\def\ps{\psi}
\def\Ps{\Psi}
\def\k{\kappa}
\def\l{\lambda}
\def\L{\Lambda}
\def\m{\mu}
\def\n{\nu}
\def\th{\theta}
\def\Th{\Theta}
\def\r{\rho}
\def\s{\sigma}
\def\S{\Sigma}
\def\ta{\tau}
\def\o{\omega}
\def\O{\Omega}
\def\pr{\prime}


\def\half{\frac{1}{2}}

\def\goto{\rightarrow}

\def\na{\nabla}
\def\grad{\nabla}
\def\curl{\nabla\times}
\def\div{\nabla\cdot}
\def\pa{\partial}
\def\fr{\frac}

\def\bra{\left\langle}
\def\ket{\right\rangle}
\def\lb{\left[}
\def\lc{\left\{}
\def\ls{\left(}
\def\lp{\left.}
\def\rp{\right.}
\def\rb{\right]}
\def\rc{\right\}}
\def\rs{\right)}

\def\vac#1{\mid #1 \rangle}


\def\td#1{\tilde{#1}}
\def\check{ \maltese {\bf Check!}}


\def\Tr{{\rm Tr}\,}
\def\det{{\rm det}}


\def\bc#1{\nnindent {\bf $\bullet$ #1} \\ }
\def\ch {$<Check!>$ }
\def\ss {\vspace{1.5cm}}
\def\text#1{{\rm #1}}
\def\Id{\mathds{1}}

\begin{titlepage}

\hfill\parbox{5cm} {APCTP Pre2016-016}

\vspace{25mm}

\begin{center}
{\Large \bf Meson's Correlation Functions in a Nuclear Medium }

\vskip 1. cm
  {Chanyong Park$^{a,b}$\footnote{e-mail : chanyong.park@apctp.org}}

\vskip 0.5cm

{\it $^a\,$ Asia Pacific Center for Theoretical Physics, Pohang, 790-784, Korea } \\
{\it $^b\,$ Department of Physics, Postech, Pohang, 790-784, Korea }\\

\end{center}

\thispagestyle{empty}

\vskip2cm


\centerline{\bf ABSTRACT} \vskip 4mm

We investigate meson's spectrum, decay constant and form factor in a nuclear medium through holographic two- and three-point correlation functions. To describe a nuclear medium composed of protons and neutrons, we consider a hard wall model on the thermal charged AdS geometry and show that due to the isospin interaction with a nuclear medium, there exist splittings of the meson's spectrum, decay constant and form factor relying on the isospin charge. In addition, we show that the $\r$-meson's form factor describing an interaction with pseudoscalar fluctuation decreases when the nuclear density increases, while the interaction with a longitudinal part of an axial vector meson increases.   

\vspace{1cm}

\vspace{2cm}


\end{titlepage}

\renewcommand{\thefootnote}{\arabic{footnote}}
\setcounter{footnote}{0}



\section{Introduction}

For describing strongly interacting nuclear physics in a nuclear medium, it is important in order to figure out low energy physics of QCD. However, since the traditional perturbation method of the quantum field theory (QFT) does not work in a strong coupling regime, we need a new paradigm or mathematical technique. In a medium, moreover, a numerical method called the lattice QCD suffers from a sign problem. In this situation, the AdS/CFT correspondence \cite{Maldacena:1997re} can shed light on accounting for a strongly interacting nuclear medium. 

Recently using the AdS/CFT correspondence, spectra of mesons and nucleons in the vaccum have been studied in various holographic models, bottom-up and top-down models \cite{Polchinski:2001tt}-\cite{Domokos:2007kt}. These works were further generalized to the medium case \cite{Lee:2009bya,Park:2011qq}. In the hard wall model, a thermal charged AdS (tcAdS) geometry has been used as the gravity dual of a nuclear medium \cite{Park:2011zp}-\cite{Adam:2015rna}. Though it contains a singularity at the center, the IR cutoff of the hard wall model prevents all physical quantities from approaching to this singularity. Therefore, the tcAdS space is safe at least in the hard wall model. 

A gauge field fluctuation on this background geometry is dual to a vector operator with a conformal dimension $3$, so it can be identified with a quark current \cite{Erlich:2005qh}. Especially, its time component corresponds to the quark density. The tcAdS geometry can be regarded as the dual of a confining phase at low temperature due to the IR cutoff, while the deconfining phase is described by a charged black hole geometry \cite{Lee:2009bya}. The Hawking-Page transition between these two geometries can be identified with the deconfinement phase transition of the dual QFT \cite{Herzog:2006ra}. In the confining phase, the fundamental excitations are not quarks but hadrons because of the confinement. Thus, the holographic model defined on the tcAdS space would be helpful to understand hadronic physics in a nuclear medium like a neutron star and nuclear physics in the strong coupling regime. In \cite{Park:2011zp}, various meson's dispersion relations and the decay constants in a nuclear medium have been studied, which was further generalized to the nucleon's cases \cite{Lee:2014xda}. These quantities related to two-point correlation functions represent hadron's spectrum and its stability. Another important thing to understand nuclear physics is a form factor describing hadron's interaction which is generally associated with a three-point correlation function \cite{Grigoryan:2007vg}-\cite{Brodsky:2011xx}. In this work, we will clarify the meson's holographic two- and three-point functions in a nuclear medium and study how they rely on the nuclear density and meson's isospin charge.

The rest of this paper is organized as follows: After briefly summarizing the holographic dual of the nuclear medium in Sec. 2, we investigate two-point functions of mesons in Sec. 3. We show that a specific Dirichlet boundary condition at the asymptotic boundary is related to the pole structure of the meson's two-point function. Using the Sturm-Liouville theorem we obtain two-point functions of vector and axial-vector mesons with a well defined decay constants. We also discuss about pion's decay constants which are not clearly identified in a nuclear medium. In Sec. 4, we study the $\r$-meson's form factors which are split because of the isospin interaction with the nuclear medium. We finish this work with some concluding remarks in Sec. 5.


\section{Holographic description for a nuclear medium}

In general, a nuclear medium is a complex system including various nucleons and mesons together with their nontrivial interactions. Although it is formidable to construct an exact gravity dual of such a system, the AdS/CFT correspondence is still useful to understand qualitative features of a strongly interacting system. In a holographic set-up, a nuclear medium can be imitated by an asymptotic AdS geometry including corresponding dual bulk fields. Since a global symmetry of a QFT maps to a local symmetry in the gravity dual, we should take into account an asymptotic AdS geometry involving at least $U(2)$ local gauge fields in order to describe a nuclear medium with two flavor charges. The tcAdS geometry has been regarded as such a dual for a nuclear medium \cite{Lee:2009bya,Park:2011zp}. The action governing the tcAdS space is given by (here we follow conventions in \cite{Park:2011zp})
\bea
S =\int d^{5}x \sqrt{-G}  \left[
\frac{1}{2\kappa ^{2}}\left( \mathcal{R}-2\Lambda \right)
- \frac{1}{4g^{2}}  \ls {F}^{(L)}_{MN} {F}^{(L)MN} + {F}^{(R)}_{MN} {F}^{(R)MN} \rs \rp  \lp +\left| D_M\Phi \right|^2 +m^2\left|\Phi \right|^2 \rb \label{2} ,
\eea
where $\Lambda =-6/R^{2}$ is a cosmological constant and the gauge field strengths of $U(2)_L$ and $U(2)_R$ are 
\bea
F^{(L)}_{MN} &=& \partial _{M} L_{N} -\partial _{N} L_{M} - i \lb L_M, L_N\rb   \quad {\rm with} \ L_M \equiv L_M^a T^a ,\nn
F^{(R)}_{MN} &=& \partial _{M} R_{N} -\partial _{N} R_{M} - i \lb  R_M,  R_N \rb \quad {\rm with} \ R_M \equiv R_M^a T^a ,
\eea
where $T^a$ indicates $U(2)$ generators. Above we considered two kinds of flavor group, $U(2)_L \times U(2)_R$,
in order to describe parity of hadrons and introduced a massive complex scalar field with $m^2 = -3/R^2$ whose modulus
represents the chiral condensate \cite{Erlich:2005qh}. A covariant derivative of $\Ph$ is defined as
\be
D_M \Ph = \pa_M \Ph - i L_M \Ph + i \Ph R_M .
\ee

For describing a nuclear matter, it is sufficient to consider only the Cartan subgroups, $U(1)_L^2 \times U(1)_R^2$, because a nuclear medium can be classified by their quantum numbers. If turning on time component gauge fields,  $L^a_t$ and $R^a_t$ with $a=0$  or $3$, they represent the quark number and isospin charge densities of the nuclear medium. Rewriting them in terms of symmetric and antisymmetric combinations the symmetric one represents a parity even state, while the antisymmetric one describes a parity odd state. The lowest parity even state is identified with proton or neutron relying on the isospin charge. Since the parity even state has lower energy than the parity odd state, the lowest parity even states are usually the main ingredients of a nuclear medium at sufficiently low energy scale. In the dual gravity, it can be accomplished by taking $L^a_t = R^a_t$, which picks up the parity even state only.  Furthermore, since fundamental excitations in the confining phase are not quarks but nucleons, we need to rewrite quark's quantities in terms of nucleon's ones. In the hard wall model, the confining phase is realized by an IR cutoff denoted by $z_{IR}$ and the tcAdS geometry with two flavor charges appears as a solution
(see \cite{Lee:2009bya,Park:2011zp} for details) 
\be		\la{back:geom}
ds^2 = \fr{R^2}{z^2} \ls - f(z) dt^2 + \fr{1}{f(z)} dz^2 + d \vec{x}^2  \rs ,
\ee
with
\bea
f(z) &=& 1 + \fr{3 Q^2 \k^2}{g^2 R^2} z^6 + \fr{D^2 \k^2}{3 g^2 R^2} z^6 , \nn
V^0_t &=& \fr{Q}{\sqrt{2}} \ls 2 z_{IR}^2 - 3 z^2 \rs, \nn
V^3_t &=& \fr{D}{3 \sqrt{2}} \ls 2 z_{IR}^2 - 3 z^2 \rs 
\eea
where $V^a_t = - \ls L^a_t + R^a_t\rs/2 $. Above $Q = Q_P + Q_N$ and $D = Q_P - Q_N \equiv \a Q$ denote the total nucleon number density and density difference between proton and neutron, where $Q_p$ and $Q_N$ indicate the number of proton and neutron respectively. 
On this tcAdS space, the deconfinement phase transition, the symmetry energy and meson's spectra with a $SU(2)$ flavor charge have been studied in \cite{Park:2011zp}.

Before closing this section, it is worth noting that when a complex scalar field $\Ph$ has a negative mass, $m^2 = - 3/R^2$, it is dual to the chiral condensate. More precisely, parameterizing the complex scalar field as
\be  \la{res:scalarmodulus}
\Ph =  \ph \Id \ e^{i \sqrt{2} \pi} ,
\ee
$\ph$ describes the chiral condensate, while the $SU(2)$ fluctuation, $\pi$, corresponds to pseudoscalar meson, the so-called pion. In the tcAdS space, the modulus satisfying the equation of motion is given by  \cite{Lee:2009bya,Park:2011zp}
\be
\ph (z) = m_q \ z \ _2 F_1 \ls \frac{1}{6} , \half , \frac{2}{3}, - \frac{\ls D^2 + 9 Q^2 \rs
z^6 }{3 \ N_c} \rs
 + \s \ z^3 \ _2 F_1 \ls \half, \frac{5}{6},\frac{4}{3},  - \frac{\ls D^2 + 9 Q^2 \rs
z^6 }{3 \ N_c}\rs ,
\ee
where $m_q$ and $\s$ denotes the current quark mass and chiral condensate respectively and $N_c$ denotes the rank of the gauge group. The gravitational backreaction of the scalar field slightly changes the background geometry and gives rise to an $1/N_c$ correction \cite{Lee:2010dh}. From now on, we ignore such a correction as usually done in the hard wall model.

\section{Meson's two-point function in a nuclear medium}

Now, let us take into account fluctuations of the gauge fields, $L^a_M  \to L^a_M + l^a_M$ and $R^a_M  \to R^a_M + r^a_M$, which can be reinterpreted as various mesons of the dual field theory.
Focusing on the spatial components by setting $l^a_t=r^a_t=0$ with the axial gauge $l^a_z=r^a_z=0$, the left and right gauge fluctuations are decomposed into vector and axial-vector fluctuations 
\be			
l^a_m = \fr{1}{\sqrt{2}} \ls v^a_m + a^a_m \rs \quad  {\rm and } \quad
r^a_m = \fr{1}{\sqrt{2}} \ls v^a_m - a^a_m \rs .
\ee
In order to interpret them as mesons, we need to redefine the vector field as
\be		\la{def:vector}
v^a_m = \r^0_m   \quad  , \quad  v^1_m =  \fr{1}{\sqrt{2}}  \ls \r^{+}_m + \r^{-}_m  \rs
\quad  {\rm and } \quad  v^2_m =  \fr{i}{\sqrt{2}}  \ls \r^{+}_m - \r^{-}_m  \rs .
\ee
Hereafter, we denote a SU(2) flavor charge as a superscript and a spatial coordinate as a subscript. Similarly, axial-vector and pion are also rewritten as the form representing the SU(2) charge manifestly
\bea        \la{def:avector}
a^a_m = a^0_m   \quad  , \quad  a^1_m &=&  \fr{1}{\sqrt{2}}  \ls a^{+}_m + a^{-}_m  \rs
\quad  , \quad  a^2_m =  \fr{i}{\sqrt{2}}  \ls a^{+}_m - a^{-}_m  \rs , \nn
\pi^3 = \pi^0   \quad  , \quad  \pi^1 &=&  \fr{1}{\sqrt{2}}  \ls \pi^{+} + \pi^{-} \rs
\quad  {\rm and } \quad  \pi^2 =  \fr{i}{\sqrt{2}}  \ls \pi^{+} - \pi^{-}  \rs .
\eea
In general, the axial-vector field couples to the pseudoscalar field. In order to remove their coupling, we rewrite the axial-vector as
\be        \la{def:agaugefix}
a^a_m = \bar{a}^a_m + \pa_m \chi^a ,
\ee
and impose the gauge fixing, $0 = \pa^m \bar{a}^a_m $. Then, the longitudinal mode of the axial-vector represents another pseudoscalar field.

In the vacuum dual to the thermal AdS geometry, meson's masses in the confining phase and the deconfinement phase transition have been studied \cite{Erlich:2005qh}. Mesons in the vacuum are degenerate and their energy satisfies a relativistic dispersion relation, $m = \sqrt{\o^2 -  |\vec{p}|^2}$, due to the Lorentz symmetry of the boundary space. In more details, if the IR cutoff is fixed, the meson's dispersion relation is uniquely determined by solving the equation of motion of the dual bulk field with two appropriate boundary conditions, a Neumann boundary condition at the IR cutoff and a Dirichlet boundary condition at the asymptotic boundary. The same dispersion relation naturally appears as a pole of a meson's two-point function. Since the analytic solution of $\r$-meson in the thermal AdS space has been known as Bessel functions \cite{Erlich:2005qh}, its exact two-point function is also known. By using the Kneser-Sommerfeld expansion (see \cite{Grigoryan:2007vg} and references therein), the two-point function of the $\r$-meson's current, which corresponds to the dual operator of $\r_\m^a$,  can be written as the covariant form
\be		\la{res:two000}
i \int \fr{d^4 x}{(2 \pi)^4} \ e^{ipx} \bra J^{a}_{\m} (x) \ J^{b}_{\n} (0) \ket =
\d^{ab}\  P_{\m\n}  \  \Pi (-p^2)  ,
\ee
where $P_{\m\n} = \fr{p_{\m} p_{\n}}{p^2} - \eta_{\m\n} $ and $\Pi (-p^2)$ denotes a two-point function in the momentum space. Here $\eta_{\m\n}$ indicates the Minkowski metric and $-p^2 = - p^{\m} p_{\m}= \o^2 - |\vec{p}|^2$. The two-point function in the momentum space is given by
\be			\la{res:twopointfninvac}
\Pi (-p^2)  = \sum_{n} \fr{  \ls D_{(n)}^a \rs^2 }{ - p ^2 -  m_{(n)}^2 }   .
\ee
where $D_{(n)}^a$  and $m_{(n)}$ are the decay constant  and mass of the $n$-th $\r$-meson resonance respectively. This result shows that the relativistic dispersion relation appears as a pole at the $n$-th resonance energy, $\o =\sqrt{ |\vec{p}|^2 + m_{(n)}^2}$. 

When a particle moves in a medium, its dispersion relation is affected from the interaction with a medium. Due to this reason, one can easily expect that the meson's dispersion relation in a nuclear medium is deviated from a relativistic one \cite{Park:2011zp}. To understand the medium effect more precisely, let us consider the following Fourier mode expansion
\be
\r^0_{m} (z,t,\vec{x})
= \int \frac{d^4 p}{(2 \pi)^4} \ \r^{0} (z,\o,\vec{p}) \  \td{\r}_m^{0} (t,\vec{x};\o,\vec{p})
\ee
where the mode function $\r^{0}$ is governed by the linearized differential equation \cite{Lee:2009bya}
\be		\la{eq:neutralrho}
0 = \pa_z \ls \fr{f(z)}{z} \pa_z   \r^{0}  \rs + \fr{1}{z} \ls \fr{\o^2}{f(z)} -  |\vec{p}|^2 \rs  \r^{0}  ,
\ee
and  $\td{\r}_m^{0} (t,\vec{x};\o,\vec{p})$ represents a neutral $\r$-meson. Now, let us normalize the kinetic part of vector meson to be  \cite{Erlich:2005qh,Park:2011zp}
\be
\int d^4 x \int_0^{z_{IR}} dz \ \sqrt{- G} \ G^{tt} G^{mn}  \ \pa_t \r^0_m \ \pa_t \r^0_n = \int d^4 x \sqrt{- \et} \ \o^2 \ls \td{\r}^0_m \rs^2 .
\ee
Then, it leads to the following normalization of the mode function
\be			\la{res:normvector}
1 = \int_0^{z_{IR}} dz \  \fr{ \ls \r^{0} \rs^2 }{z  f(z)} .
\ee
The translation symmetry at the boundary relates $\td{\r}_m^{0} $ to a plane wave, $\td{\r}_m^{0} = \e_m^0 e^{- i \ls  \o t - \vec{p} \cdot \vec{x} \rs}$, with a polarization vector $\e_m^0$. In this case, the rest mass of $\r$-meson is given by the energy at $|\vec{p}|=0$. In the nuclear and isospin medium the medium effect on the rest mass has been studied in \cite{Lee:2014xda,Lee:2010dh}.

In the asymptotic region ($Q^2 z^6 \ll 1$), the mode function allows the following perturbative solution
\be			\la{res:perexp0}
\r^{0}  = c^0 + \td{c}^0  \ z^2  + \cdots  ,
\ee
where $c^0 $ and $\td{c}^0$ are two independent integral constants. In order to fix them, we impose the following boundary conditions  
\be		\la{con:rhoboundarycond}
\lim_{z \to 0} \ \r^0 = c^0   \quad  {\rm and}  \quad  \lim_{z \to z_{IR}} \
\pa_z \r^0 = 0 .
\ee
Since the equation of motion in \eq{eq:neutralrho} includes several free parameters, $Q$, $D$, $\o$ and $p$, the boundary value of $\r$-meson can also have a nontrivial dependence on them. When the value of $c^0$ is given, the free parameters have to satisfy a specific relation corresponding to the dispersion relation.

Let us further investigate two-point functions in the nuclear medium. Since the analytic solution of \eq{eq:neutralrho} is not known, one can not directly applied to the Kneser-Sommerfeld  expansion unlike the vacuum case. In spite of this, one can find a similar two-point function form even in the nuclear medium. To see this, we first calculate the on-shell gravity action in the momentum space
\be              
S_B \equiv   - \int_{\pa {\cal M}} \fr{d^4 p }{(2 \pi)^4} {\cal L}_B= - \frac{2}{g_5^2} \int_{\pa {\cal M}}  \fr{d^4 p }{(2 \pi)^4} \   c^{0}  \ \td{c}^{0} ,
\ee
where the minus sign appears because the direction of the normal vector is opposite to the $z$-direction. The two-point function of the neutral $\r$-meson reads from the on-shell action,
\be 	\la{res:twopointfunrho}
\Pi^0 (-p^2) \equiv
 - \lim_{z \to 0}   \half \fr{\pa^2 {\cal L}_B }{\pa c^0   \ \pa c^0} 
=  - \frac{2  }{g_5^2} \fr{\pa \td{c}^0}{\pa c^0}   .
\ee
Due to the Neumann boundary condition at the IR cutoff, $\td{c}^0$ should be related to $c^0$. 
Since $\r^0$ is the solution of the linearized homogeneous differential equation  a new function, $y=\l \r^0$, scaled by a constant $\l$ still remains as a solution. This scaling allows us to write $\td{c}^0 = N c^0$ where $N$ is independent of $c^0$. As a result, the scaling independent two-point function becomes in terms of $c^0$ and $\td{c}^0$
\be
\Pi^0 (-p^2)  = - \fr{2}{g_5^2} N =  - \fr{2}{g_5^2} \fr{\td{c}^0}{c^0} .
\ee 
In general, resonances appear as a pole of the two-point function. Assuming that $\td{c}_0$ corresponding to the decay constant is finite, in the above two-point function the resonance can be obtained by imposing the additional condition, $c^0 =0$.
Near the $n$-th resonance energy, the boundary value of the mode function can be expanded into
\be  \la{res:resonancecoeff}
 c^0   =   \fr{ \o^2 - \ls \S_{(n)}^0 \rs^2  }{ \cal N }
\lb 1 + {\cal O} \ls  \o^2 - \ls \S_{(n)}^0 \rs^2 \rs  \rb ,
\ee
where ${\cal N}$ is introduced as a normalization factor and the medium effect is encoded into the $n$-th resonance energy, $\S_{(n)}^0$. 
This relation shows that the boundary value of $\r^0$ out of the exact resonance energy does not vanish, so the resonances appear at discrete energy values. 

Taking the following normalization constant for consistency
\be
{\cal N} = -  g_5  {D^0_{(n)}} ,
\ee
the decay constant is determined from the $n$-th resonance, ${\r}_{(n)}^0$, satisfying ${\r}^0=0$ at $z=0$
\be
D^0_{(n)}  = \lim_{z \to 0} \ \fr{1}{g_5} \fr{\pa_z {\r}_{(n)}^0}{ z}  = 2 \td{c}_{(n)}^0 ,
\ee
where $\td{c}_{(n)}^0$ indicates the value of $\td{c}^0$ in ${\r}_{(n)}^0$.
In general, the dominant contribution to the two-point function comes from the resonance with the energy, $\o=\S_{(n)}^0$,
\be		\la{res:two000mode}
\Pi^0 (-p^2)
=  \fr{\ls {D^0_{(n)} }\rs^2  }{\o^2 - \ls \S^0_{(n)}  \rs^2 }   .
\ee
Following the Sturm-Liouville theorem, a general two-point function can be rewritten as the sum of resonance's two-point functions 
\be		\la{res:two000mode1}
\Pi^0 (-p^2)
=  \sum_n  \fr{\ls {D^0_{(n)} }\rs^2  }{\o^2 - \ls \S^0_{(n)}  \rs^2 }   .
\ee
The two-point function of $\r^0$-meson in the nuclear medium has a similar form to that in the vacuum. Its pole represents a dispersion relation of the $n$-th resonance. However, due to the interaction with a medium, the dispersion relation in a nuclear medium cannot be written as a relativistic form. Note that there exists an alternative way to derive \eq{res:two000mode1} following the prescriptions used in \cite{Erlich:2005qh}, which also gives rise to the exact same result.

Now, let us move to charged $\r$-mesons whose mode function satisfies the following equation \cite{Park:2011zp}
\be
0 = \partial_z\left(\frac{f(z)}{z}\partial_z{\rho}^\pm \right) +\frac{1}{z}\left[ \fr{1}{f(z)} \left(w _\pm\mp\bar{V}^3_t\right)^2 - p^2\right]{\rho}^\pm .
\ee
Near the asymptotic region, it allows the perturbative solution
\be
\r^{\pm}   = c^{\pm}  +  \td{c}^{\pm}   \ z^2  + \cdots   ,
\ee
where  two integral constants, $c^{\pm}$ and $\td{c}^{\pm}$, become near a resonance energy
\bea
c^{\pm}  &=&
-  \fr{    \o^2 - \ls \S^{\pm}_{(n)}  \rs^2}{   g_5  {D^{\pm}_{(n)}} } , \nn
\td{c}^{\pm} &=& \fr{D^{\pm}_{(n)}}{2}     = \lim_{z \to 0} \ \fr{1}{2 g_5} \fr{\pa_z \r^{\pm}}{ z}  .
\eea
Then, the holographic two-point function is reduced to
\be
 \Pi^{\pm} (-p^2)  = \sum_{n}   \fr{ \ls {D^{\pm}_{(n)}} \rs^2 }{ \o^2 - \ls \S^{\pm}_{(n)} \rs^2} ,
\ee
with the following decay constants
\be
D^{\pm}_{(n)}  = \lim_{z \to 0} \ \fr{1}{g_5} \fr{\pa_z {\r}_{(n)}^{\pm}}{ z}   .
\ee
In a nuclear medium, calculating the two-point function of axial-vector mesons is parallel to the vector meson case, so we skip the details. 

Lastly, let us discuss a two-point function of pions. Using the following Fourier mode expansion
\be
\pi^a (z,t,\vec{x})
= \int \frac{d^4 p}{(2 \pi)^4} \ \pi^a (z,\o,\vec{p}) \  \td{\pi}^a  (t,\vec{x};\o,\vec{p}) ,
\ee
pions in the nuclear medium are governed by \cite{Park:2011zp} 
\bea
\frac{z^3 f(z)}{g^2\phi^2} \partial_z\left(\frac{g^2\phi^2 f(z)}{z^3}\partial_z {\pi}^0\right) 
&=&  \left(w^2_0-f(z) |\vec{p}|^2\right)\left( {\chi}^0 -  {\pi}^0\right)  ,\nonumber\\
 \frac{z^3 f(z)} {4g^2\phi^2}\partial_z\left(\frac{w_0^2-f(z) |\vec{p}|^2}{z}\partial_z {\chi}^0\right)
&=&\left(w^2_0- f(z)  |\vec{p}|^2\right)\left({\chi}^0 -  {\pi}^0\right)   ,\nonumber\\
 \frac{z^3 f(z)} {g^2\phi^2}\partial_z\left(\frac{g^2\phi^2 f(z)}{z^3}\partial_z {\pi}^\pm\right)  
 &=& \left( w_\pm^2 \mp {V}^3_t w_\pm - f(z) |\vec{p}|^2\right) {\chi}^\pm - \left( \left(w_\pm \mp{V}^3_t\right)^2 -f(z) |\vec{p}|^2\right) {\pi}^\pm   ,\nonumber\\
\frac{z^3 f(z)} {4g^2\phi^2}\partial_z\left(\frac{w_\pm^2-f(z) |\vec{p}|^2}{z}\partial_z {\chi}^\pm\right)  
&=& \left(w^2_\pm-f(z) |\vec{p}|^2 +\frac{ \ls {V}^3_t\rs^2 z^2}{4g^2\phi^2}|\vec{p}|^2\right) {\chi}^\pm \nn
&& -\left(w^2_\pm \mp{V}^3_tw_\pm- f(z)  |\vec{p}|^2\right) {\pi}^\pm .
\eea
Unlike the vector and axial vector mesons, pions are described by two fields, $\chi^a$ and $\pi^a$. $\pi^a$ corresponds to a fluctuation of a pseudoscalar field, while $\chi^a$ comes from the longitudinal mode of the axial-vector field. In the chiral limit ($m_{\pi}=0$), the pion decay constant in the vacuum is defined as $ \bra 0  | J^{a}_m | \pi \ket = i D^a_{\pi} p_m $ where $J^{a}_m$ indicates an axial-vector current \cite{Erlich:2005qh}. Due to this relation, the decay constant of pion is usually evaluated from the two-point function of $a_1$-meson. In the holographic point of view, this is reasonable because $\pi= 0$ becomes a solution in the chiral limit and at the same time $\chi^a$ satisfies the same equation of axial-vector meson. In a nuclear medium, however, this is not the case because of a nontrivial isospin interaction with the background matter.  Even in the chiral limit, $\pi^{\pm} =0$ is not a solution and the equation for $\chi^{\pm}$ does not reduce to that of axial-vector meson. These facts imply that we need a new method to define pion's decay constant in the nuclear medium. We leave this issue as a future work. Although the pion's decay constant is not well defined in the nuclear medium \cite{Sheel:1994ak}, two-point functions of $\pi^a$ and $\chi^a$ should have the same dispersion relation. In the nuclear medium, we depict energies and decay constant of $\r$-mesons and pions with a specific momentum preserving SO(2) rotation symmetry (see Fig. 1 and Fig. 2). Since $\pa_m \chi^a$ and $\r^a_m$ have the same origin as shown in \eq{def:vector}, \eq{def:avector} and \eq{def:agaugefix}, one can use the similar normalization used in \eq{res:normvector} 
\be
1 = \int_0^{z_{IR}} dz \ \fr{\ls \chi^a \rs^2}{z f(z)} .
\ee
However, the different origin of the kinetic term for $\pi^a$ requires a different normalization    \cite{Park:2011zp}   
\be
\int d^4 x \int_0^{z_{IR}} dz \ \sqrt{- G} \ G^{tt}   \ \pa_t \pi^a \ \pa_t \pi^a = \int d^4 x \sqrt{- \et} \ \o^2 \ls \td{\pi}^a\rs^2 ,
\ee
which yields the following normalization  
\be
1 = \int_0^{z_{IR}} dz \ \fr{\ls \pi^a \rs^2}{z^3 f(z)}  .
\ee

\begin{figure}
\begin{center}
\vspace{0cm}
\hspace{-0.5cm}
\subfigure[$\r$-mesons]{ \includegraphics[angle=0,width=0.45\textwidth]{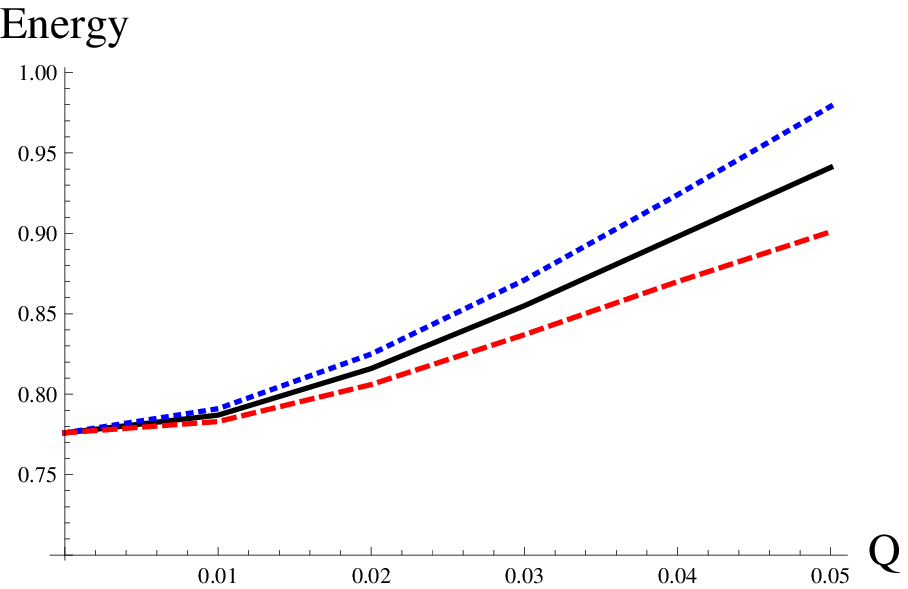}}
\hspace{1cm}
\subfigure[pions]{ \includegraphics[angle=0,width=0.45\textwidth]{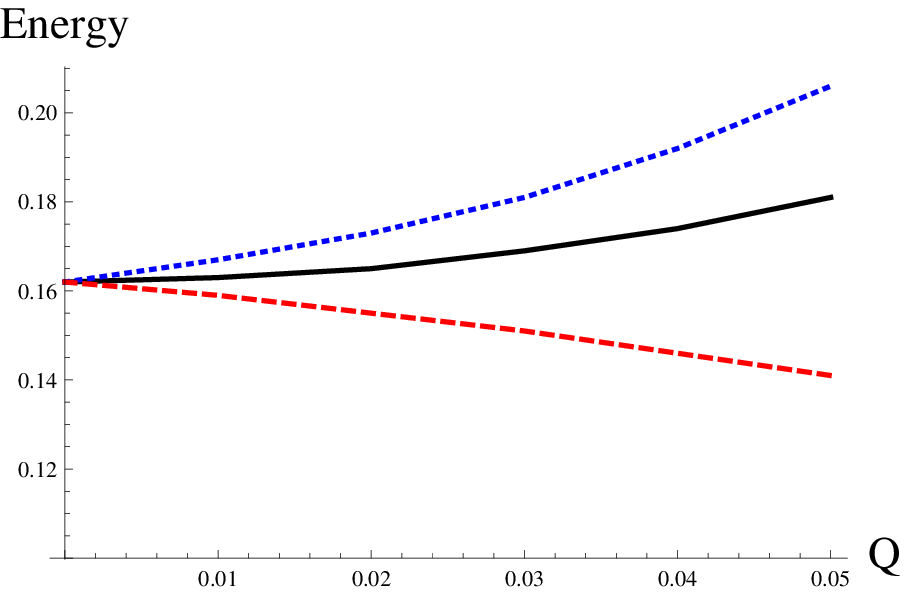}}
\vspace{-0cm}
\caption{\small The first resonance energies of mesons, $\S^a_{(1)}$, in the nuclear medium. When $\a=-1/2$, the energies of $\r$-mesons with $\vec{k} = \lc 0, 0, 0.01 \rc$ is depicted in (a). pions with $\vec{k} = \lc 0.04, 0.04, 0.01 \rc$ show their energies in (b). Above a solid (black) curves describe the energy of a neutral mesons,
whereas dotted and dashed curves represent those of negatively and positively charged mesons respectively. }
\label{number}
\end{center}
\end{figure}

In general, neutron in a nuclear medium is more stable than proton because of the electric interaction of protons. Therefore, the neutron dominant case ($\a < 0$) is more physical. When $\a=-1/2$, the nuclear medium we considered is composed of $75 \%$ neutrons and $25 \%$ protons. In this case, the energy of a negatively charged meson is usually larger than that of a positively charged meson because of the isospin interaction with the nuclear medium as shown in Fig. 1. In addition, their energy always increases when the nuclear medium density increases. Similarly, the decay constant of $\r$-mesons also increases as shown in Fig. 2. In general, there also exists the splitting of the decay constant in the nuclear medium  \cite{Lee:2014xda}. However, since we have taken into account only a small $Q$ range in Fig. 2, the splitting of the decay constant is negligible.

\section{ $\r$-meson's form factor}

Following the AdS/CFT correspondence, bosonic bulk fluctuations map to various mesons of the dual field theory. Their interactions can be studied from higher point correlation functions. On the gravity side, those $n$-point functions are described by interactions of dual bulk fluctuations. Since pions and $\r$-mesons have relatively smaller mass than others, they are usually dominant in the low energy physics and play a crucial role in understanding the strongly interacting nuclear matter. In this section, we will focus on the form factor of $\r$-mesons in the nuclear medium. This work can be easily extended to other massive mesons like $a_1$-meson and higher resonances.

\begin{figure}
\begin{center}
\vspace{0cm}
\hspace{-0.5cm}
\subfigure{ \includegraphics[angle=0,width=0.45\textwidth]{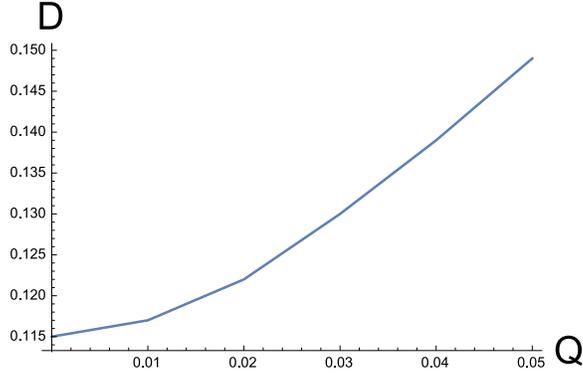}}
\caption{\small Decay constant of $\r$-mesons in the nuclear medium with $0< Q \le 0.05$ and $\vec{k} = \lc 0, 0, 0.01 \rc$. Note that in the parameter range we used the splitting of decay constants is negligible.}
\label{number}
\end{center}
\end{figure}

The form factor can be read from the three-point function, so that it is required to expand the bulk action into cubic order. Terms related to $\r$-meson at cubic order are given by
\bea
S_{\r } &=& \int d^5 x \sqrt{-g} \
\lb  - \fr{1}{4 g^2}  \e^{abc}    \lc  \ls \pa_m l^a_n - \pa_n l^a_m \rs  l^{bm} l^{cn} + \ls \pa_m r^a_n - \pa_n r^a_m \rs  r^{bm} r^{cn} \rc \rp  \nn
&& \lp - \ph^2  \e^{abc}  \lc \pa_m \pi^a \ls l^{bm} + r^{bm} \rs \pi^c  -
 \pa_m \pi^a \pi^b \ls l^{cm} + r^{cm} \rs  \rc \rb  ,
\eea
where $l^a_m$ and $r^a_m$ are bulk gauge field fluctuations of $SU(2)_L$ and  $SU(2)_R$ sectors, respectively. 
Using the definitions in \eq{def:vector}, \eq{def:avector} and \eq{def:agaugefix}, the cubic order action governing the interaction between $\r$-meson and pions reduces to
\bea
S_{\r } &=& \int d^5 x \sqrt{-g} \ \lb  -\fr{i}{4 \sqrt{2} g^2} \lc
 - \ls \pa_m \r^0_n -  \pa_n \r^0_m  \rs \ls \pa^m \chi^{+} \pa^n \chi^-  -  \pa^m \chi^- \pa^n \chi^+ \rs \rp
\fr{}{}\rp \nn
&& \qquad \qquad  \qquad  \qquad  \quad  \quad
+ \ls \pa_m \r^-_n -  \pa_n \r^-_m  \rs \ls \pa^m \chi^{+} \pa^n \chi^0 -  \pa^m \chi^{0} \pa^n \chi^+ \rs  \nn
&& \qquad \qquad  \qquad  \qquad  \quad  \quad \lp
- \ls \pa_m \r^+_n -  \pa_n \r^+_m  \rs \ls \pa^m \chi^- \pa^n \chi^0 -  \pa^m \chi^{0} \pa^n \chi^- \rs  \rc \nn
&& \qquad \qquad \qquad  - i 2 \sqrt{2} \ph^2 \lc \r^0_m \ls \pi^+ \pa^m \pi^- - \pa^m \pi^+ \pi^-\rs
+ \r^-_m \ls \pi^+ \pa^m \pi^0 - \pa^m \pi^+ \pi^0\rs \rp \nn
&& \lp \qquad \qquad  \qquad  \qquad  \quad  \quad \lp - \r^+_m \ls \pi^- \pa^m \pi^0 - \pa^m \pi^- \pi^0 \rs  \rc \rb  .
\eea
In terms of momentum, it can be further reduced to
\bea
S_{\r} &=&  - \int d^4 x  \
\lb  F_{\r\chi\chi} (0,+,-) \ \td{\r}^0_m  \td{\chi}^+  \td{\chi}^-
+  F_{\r\chi\chi} (+,0,-) \ \td{\r}^+_m  \td{\chi}^0  \td{\chi}^- \rp \nn
&& \qquad \qquad  \qquad  - F_{\r\chi\chi} (-,0,+) \ \td{\r}^-_m  \td{\chi}^0  \td{\chi}^+
+ F_{\r\pi\pi} (0,+,-)  \ \td{\r}^0_m  \td{\pi}^+  \td{\pi}^- \nn
&& \lp \qquad \qquad  \qquad  + F_{\r\pi\pi} (+,0,-) \ \td{\r}^+_m  \td{\pi}^0  \td{\pi}^-
- F_{\r\pi\pi} (-,0,+)  \ \td{\r}^-_m  \td{\pi}^0  \td{\pi}^+  \rb
\eea
with the following form factors
\bea
F_{\r\chi\chi} (a,b,c) &=& \fr{1}{2 \sqrt{2} g^2} \ \int_0^{z_{IR}}  dz  \sqrt{-g} \ k^a_n \ls k^{bn} k^{cm} - k^{cn} k^{bm} \rs
\ \r^a  \chi^b  \chi^c , \nn
F_{\r\pi\pi} (a,b,c) &=&  2 \sqrt{2} \ \int_0^{z_{IR}} dz \sqrt{-g} \  \ph^2  \ls k^{bm} - k^{cm} \rs
\ \r^a  \pi^b  \pi^c .
\eea
Recall that variables with the tilde symbol indicate mesons in the dual field theory and ones without tilde correspond to their mode functions in the dual gravity. 

In order to see the medium effect on the $\r$-meson form factor, we simplify the situation. Let us consider a $\r$-meson moving along the $z$-direction with the momentum parameterized by $\vec{k}^{a}_m = \lc 0,0, - 2 k_z \rc$, which breaks the $SO(3)$ rotational symmetry into $SO(2)$. In addition, let us consider pions preserving the remaining $SO(2)$ symmetry for simplicity. Then, the momentum conservation implies that two pions have the momenta, $\vec{k}^{b}_m  = \lc k,k,k^{b}_{z} \rc$ and $\vec{k}^c_m =\lc -k,-k,k^c_{z} \rc$, with $2 k_z = k^b_{z}+k^c_{z}$. Let us further assume that two pions have the same momentum along the $z$-direction, $k^b_{z}=k^c_{z}=k_z $. Then, the form factor we are interested in can be parameterized by $k^b$ and $k_z$ only. Finally the form factors are reduced to
\bea
F_{\r\chi\chi} (a,b,c)
&=& \fr{2 \sqrt{2}}{ g^2} \ k_z^2  k \ \int_0^{z_{IR}}  dz \sqrt{-g} \ g^{zz} g^{xx} \  \r^a  \chi^b  \chi^c   , \nn
F_{\r\pi\pi} (a,b,c)
&=& 8 \sqrt{2} \ k \ \int_0^{z_{IR}}  dz \sqrt{-g} \ \ph^2 g^{xx}  \ \r^a  \pi^b  \pi^c ,
\eea
where the SO(2) invariance is used. Note that mode functions, $\r^a $, $ \chi^a$ and $ \pi^a$, nontrivially depend on the properties of the medium and meson's momenta. In Fig. 3, we depict the $\r$-meson form factors depending on the nuclear medium properties with a fixed momenta, $k_z = 0.01$ and $k=0.04$. Intriguingly, $F_{\r\pi\pi}$ and $F_{\r\chi\chi}$ show totally different behaviors. When the nuclear density increases the form factors of $\r$ and $\pi$ decreases, while the form factors of $\r$ and $\chi$ increases.

Following the AdS/CFT correspondence, the results obtained here can be regarded as the ones including all quantum effects. In this sense, the form factor may be regarded as a non-perturbative effective coupling constant between $\r$-meson and pions in the nuclear medium. If ignoring the isospin effect, the form factors of $\r$-mesons become degenerate. In other words, there is no distinct between the form factors of neutral and
charged $\r$-mesons. Turning on the isospin effect, the split of the form factors occurs as expected (see Fig. 3). 
For $\a=-1/2$, the interaction between a negatively charged $\r$-meson and $\pi$ is stronger than that 
between a positively charged $\r$-meson and $\pi$. On the contrary, the interaction between a negatively 
charged $\r$-mesion and $\chi$ becomes weaker than that of a positive $\r$-meson.

\begin{figure}
\begin{center}
\vspace{0cm}
\hspace{-0.5cm}
\subfigure{ \includegraphics[angle=0,width=0.45\textwidth]{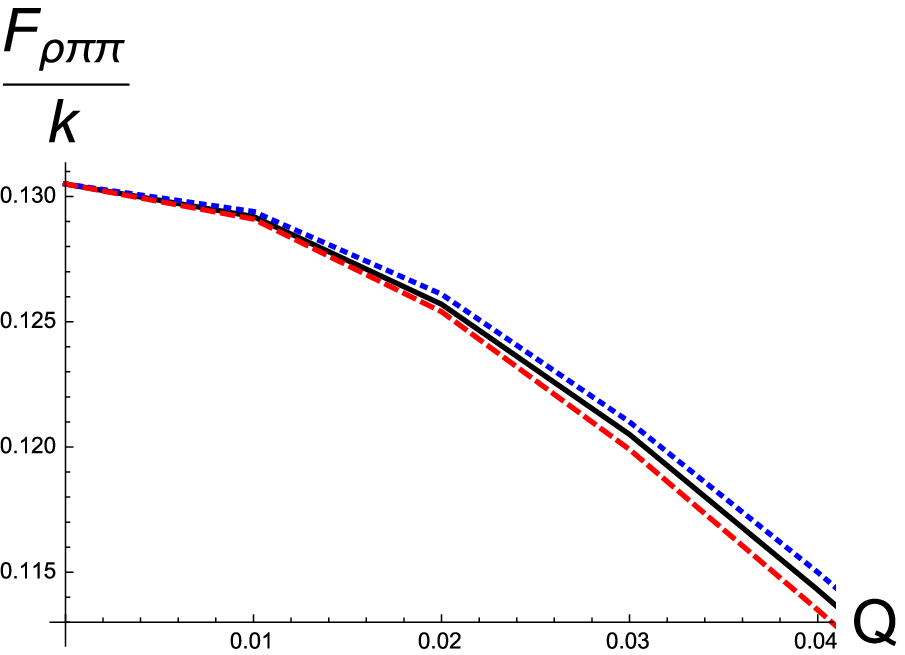}}
\hspace{1cm}
\subfigure{ \includegraphics[angle=0,width=0.45\textwidth]{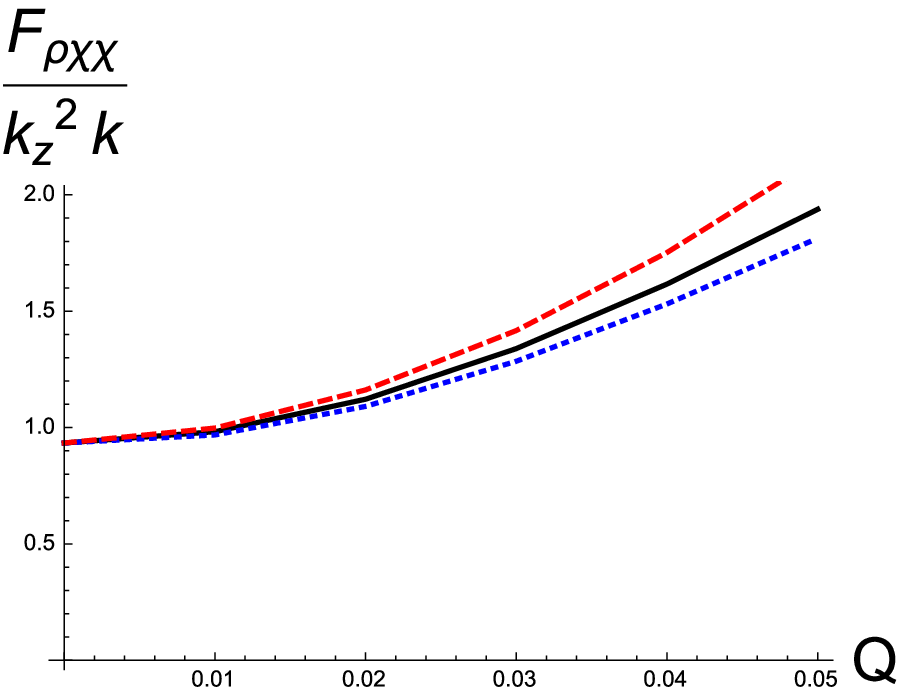}}
\vspace{-0cm}
\caption{\small  Form factors in the nucelar medium with $\a=-1/2$. Above, a solid (black) line represents the form factor
of a neutral $\r$-meson, whereas a dotted (blue) and dashed (red) curves indicate those of a negatively and  positively charged $\r$-meson respectively.  }
\label{number}
\end{center}
\end{figure}


\section{Discussion}

In this work, we have investigated meson's two-point functions and form factors in a nuclear medium. Imposing the Dirchlet boundary condition on the dual bulk field at the asymptotic boundary describes the pole structure of the meson's two-point function. Although the analytic solution of the dual bulk field is not known in the  nuclear medium, the Sturm-Liouville theorem allows us to write the vector and axial-vector meson's two-point function with a well-defined decay constant. In a vacuum, the pion's decay constant can be evaluated by that of the axial-vector meson in the chiral and zero momentum limit where axial-vector meson and pion satisfy the same equation of motion. In a nuclear medium, however, they interact with the nuclear medium differently. Due to this reason, the pion's decay constant is not well defined from the axial-vector's one even in the chiral and zero momentum limit. It would be interesting to find a new method to define the pion's decay constant in a nuclear medium. 

It has been shown that the isospin interaction of mesons with the nuclear medium causes the mass splitting depending on their isospin charges. In this work, we have also shown that $\r$-meson's form factors are also split in the nuclear medium. When the nucleon's density of the nuclear medium increases the form factor, $F_{\r\pi\pi}$, between $\r$ and $\pi$  decreases, while $F_{\r\chi\chi}$ increases.

\vspace{0.5cm}

{\bf Acknowledgement}

C. Park was supported by Basic Science Research Program through the National Research Foundation of Korea funded by the Ministry of Education (NRF-2013R1A1A2A10057490) and also by the Korea Ministry of Education, Science and Technology, Gyeongsangbuk-Do and Pohang City.


\end{document}